\documentclass[aps,prx,twocolumn,superscriptaddress]{revtex4-2}

\usepackage{times}
\usepackage{braket}
\usepackage{xcolor}
\usepackage{graphicx}
\usepackage{multirow}
\usepackage{amssymb}
\usepackage{amsmath}
\usepackage[normalem]{ulem}

\usepackage{hyperref}
\hypersetup{
    colorlinks=true,
    citecolor=blue,
}

\begin{document}

\title{Observation of quantum thermalization restricted to Hilbert space fragments}
\author{Luheng Zhao}
\affiliation{Centre for Quantum Technologies, National University of Singapore, 117543 Singapore, Singapore}
\affiliation{Duke Quantum Center, Duke University, Durham, NC 27701, USA}
\affiliation{Department of Electrical and Computer Engineering, Duke University, Durham, NC 27708, USA}
\author{Prithvi Raj Datla}
\affiliation{Department of Physics, National University of Singapore, 117542 Singapore, Singapore}
\author{Weikun Tian}
\affiliation{Centre for Quantum Technologies, National University of Singapore, 117543 Singapore, Singapore}
\author{Mohammad Mujahid Aliyu}
\affiliation{Centre for Quantum Technologies, National University of Singapore, 117543 Singapore, Singapore}
\author{Huanqian Loh}
\email[]{huanqian.loh@duke.edu}
\affiliation{Centre for Quantum Technologies, National University of Singapore, 117543 Singapore, Singapore}
\affiliation{Duke Quantum Center, Duke University, Durham, NC 27701, USA}
\affiliation{Department of Physics, National University of Singapore, 117542 Singapore, Singapore}
\affiliation{Department of Electrical and Computer Engineering, Duke University, Durham, NC 27708, USA}
\affiliation{Department of Physics, Duke University, Durham, NC 27708, USA}

\begin{abstract}

Quantum thermalization occurs in a broad class of systems from elementary particles to complex materials. Out-of-equilibrium quantum systems have long been understood to either thermalize or retain memory of their initial states, but not both. Here we achieve the first coexistence of thermalization and memory in a quantum system, where we use both Rydberg blockade and facilitation in an atom array to engineer a fragmentation of the Hilbert space into exponentially many disjointed subspaces. We find that the kinetically constrained system yields quantum many-body scars arising from the $\mathbb{Z}_{2k}$ class of initial states, which generalizes beyond the $\mathbb{Z}_{2}$ scars previously reported in other quantum systems. When bringing multiple long-range interactions into resonance, we observe quantum thermalization restricted to Hilbert space fragments, where the thermalized system retains characteristics of the initial configuration. Intriguingly, states belonging to different subspaces do not thermalize with each other even when they have the same energy. Our work challenges established ideas of quantum thermalization while experimentally resolving the longstanding tension between thermalization and memory. These results may be applied to control entanglement dynamics in quantum processors and quantum sensors.

\end{abstract}

\maketitle

\section{Introduction}
Quantum thermalization is a central concept for understanding the behavior of systems ranging from atomic nuclei to black holes \cite{zelevinsky1996quantum,berges2021qcd,zhou2022thermalization,rodriguez2013thermalization,hayden2007black,polkovnikov2011colloquium,eisert2015quantum}. Quantum thermalization leads to information loss through rapid entanglement spreading, as described by the eigenstate thermalization hypothesis (ETH) that has been postulated to describe the ergodic evolution of isolated quantum systems \cite{deutsch1991quantum,rigol2008thermalization,neill2016ergodic,kaufman2016quantum}. Conversely, the counterexamples of ETH, notably integrable systems, many-body localization, and quantum many-body scars, are associated with preserving memory of the initial state \cite{kinoshita2006quantum,nandkishore2015many,schreiber2015observation,rispoli2019quantum,kohlert2019observation,schulz2019stark,morong2021observation,bernien2017probing,turner2018weak,bluvstein2021controlling,zhang2023many}. The question of how thermalization can coexist with memory --- two phenomena long considered to be at odds with each other --- has generated much debate in the past decade and remains important for finding new ways to control entanglement propagation \cite{abanin2019colloquium,serbyn2021quantum,moudgalya2022quantum}. 

Recently, Hilbert space fragmentation has been put forth as a mechanism through which quantum systems can exhibit rich entanglement dynamics \cite{sala2020ergodicity,khemani2020localization,yang2020hilbert,rakovszky2020statistical,moudgalya2021thermalization,moudgalya2022quantum,moudgalya2022hilbert,valencia2023rydberg,doggen2021stark}. In this mechanism, the Hilbert space is shattered into exponentially many disjointed subspaces, or Krylov subspaces, each of which can be integrable or non-integrable. The fragmentation can originate from approximate conservation laws or imposed kinetic constraints \cite{scherg2021observing,kohlert2023exploring,kim2023realization}. Fragmented systems are important for understanding a host of exotic phenomena including quark confinement, high-temperature superconductivity, and fractional quantum Hall physics \cite{yang2020hilbert,batista2000quantum,moudgalya2022hilbert,moudgalya2020quantum}. 

Krylov-restricted thermalization has been conjectured to occur in fragmented systems, by which ETH would be permitted only within a given non-integrable Krylov subspace but be violated with respect to the entire Hilbert space \cite{moudgalya2021thermalization,moudgalya2022quantum}. However, since the Krylov fractures were not thoroughly understood, it was not initially obvious that Krylov-restricted thermalization would occur \cite{sala2020ergodicity}. The conjecture was later supported by numerics \cite{moudgalya2021thermalization}, although it was not clear how to translate from theoretical models to an experimental implementation. After all, in physical realizations of fragmented systems, the Hilbert space is often split into energy-conserving blocks \cite{moudgalya2022quantum}. In other words, full thermalization within an energy-conserving subspace would still be consistent with global ETH, which compares a globally entangled state against a thermal ensemble of states \textit{within a narrow energy window} \cite{rigol2008thermalization}. In fact, for fragmented models derived from kinetic constraints, it is not generally known how the Krylov subspaces would be identified \cite{khemani2020localization}. Even when the Hilbert space fractures are relatively well understood as in dipole-moment-conserving models, a recent experimental study found disagreement between the steady-state observables and theoretical predictions for thermalization within the Krylov fragment, indicating that there is no guarantee for a fragmented system to thermalize within a given subspace \cite{moudgalya2021thermalization,moudgalya2022quantum,herviou2021many, de2019dynamics, kohlert2023exploring}.

\begin{figure*}[!htbp]
    \centering
    \includegraphics[keepaspectratio,width=18cm]{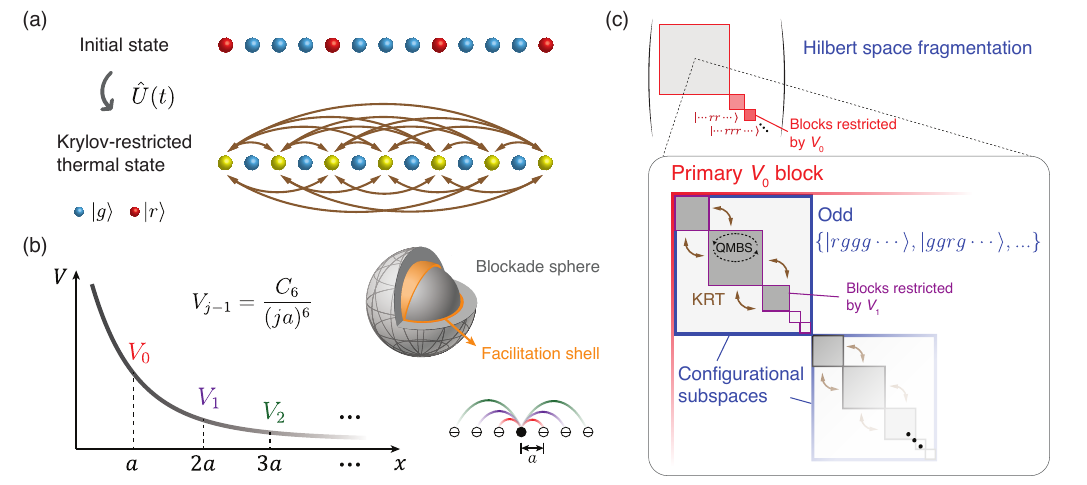}
        \caption{Krylov-restricted thermalization arising from Hilbert space fragmentation. (a) Schematic of a Rydberg atom array initialized in the $\mathbb{Z}_{4}$ ordered state and evolving into a Krylov-restricted thermal state, where atoms at odd sites thermalize with each other while atoms at even sites remain frozen. (b) The presence of a Rydberg atom can yield a blockade sphere (gray), within which Rydberg excitations are suppressed, or a facilitation shell (orange), at which Rydberg excitations are allowed. Strong kinetic constraints are engineered by facilitating different orders of the Rydberg interaction. (c) The Hilbert space in the product-state basis is first fragmented into blocks restricted by the $V_{0}$ interaction (red), where states in the primary (secondary) $V_{0}$ block have zero (one) $V_{0}$ energy term, and so on. Within each $V_{0}$ block, the Hilbert space is further fragmented into configurational subspaces (blue), e.g. the odd configurational subspace, where Rydberg excitations solely occur on odd-numbered sites of the atom chain. Within each configurational subspace, the Hilbert space is then fragmented into blocks restricted by the $V_{1}$ interaction (purple). The finer block structure iterates for higher-order interactions. Krylov-restricted thermalization (KRT) is achieved by lifting the $V_{1}$ restriction while maintaining the disconnectivity between configurational subspaces (brown arrows). }
    \label{fig:schematic}
\end{figure*}

Here we report the first observation of quantum thermalization restricted to Krylov subspaces. The Hilbert space fragmentation in our Rydberg atom array experiment arises from kinetic constraints, where we demonstrate the latter through $\mathbb{Z}_{2k}$ quantum many-body scarring, generalizing beyond the $\mathbb{Z}_{2}$ scarring previously reported in cold-atom setups \cite{bernien2017probing,bluvstein2021controlling,su2023observation}. We discover Krylov subspaces beyond energy-conserving blocks, which we identify through the configuration of Rydberg atoms [Fig.~\ref{fig:schematic}]. We observe that while thermalization proceeds in a given subspace, thermalization between energy-degenerate states belonging to different Krylov subspaces is precluded, yielding a clear violation of ETH for the entire Hilbert space. In other words, the thermalized system preserves memory of the initial configuration by storing information through the relevant Krylov subspace.

\section{Facilitated QXQ model}
In our experiment setup (Appendix~\ref{Appendix:A.1}), we use a linear chain of equally spaced $^{87}$Rb atoms ($N = 13 - 19$ atoms) with pseudospin states encoded in their ground $\ket{g}$ and Rydberg $\ket{r}$ states $(\hbar = 1)$:
\begin{equation} \label{eq:Hamiltonian}
H_{\text{Ryd}} = -\Delta \sum_{i=1}^{N} Q_{i} + \frac{\Omega}{2}\sum_{i=1}^{N} X_{i} + \sum_{j = 1}^{N-1} \sum_{i = 1}^{N-j} V_{j-1} Q_{i} Q_{i+j} \, ,
\end{equation}
where $i$ indexes the atom, $V_{j-1} = C_6/(ja)^{6}$ is the van der Waals interaction between Rydberg atoms, $a$ is the interatomic spacing, $\Delta$ is the detuning, $\Omega$ is the Rabi frequency, $Q_{i} = \ket{r_{i}}\bra{r_{i}}$ is the projector onto the Rydberg state, and $X_{i} = \ket{g_{i}}\bra{r_{i}} + \ket{r_{i}}\bra{g_{i}}$. In the $\{\ket{g}, \ket{r}\}$ basis, $Z\ket{g} = -\ket{g}$, and $Z\ket{r} = +\ket{r}$. We prepare the initial state through single-site addressing \cite{labuhn2014single} of a subset of the chain, followed by a global $\pi$ pulse for Rydberg excitation. We then turn off the single-site addressing beams and probe the subsequent dynamics by measuring the Rydberg and ground state densities after a given time evolution. 

Such a Rydberg atom array quantum simulator is often used in the Rydberg blockade regime, where the presence of a Rydberg atom prevents all other atoms within a blockade radius $R_{b}$ from being excited to the Rydberg state, regardless of their precise position within $R_{b}$. In particular, where the blockade radius only covers nearest neighbors such that $V_{0} \gg \Omega \gg V_{1}$, the constrained dynamics are captured by an effective PXP model, which has been used to explain quantum many-body scars of $\mathbb{Z}_{2}$ order (e.g.\ revivals between $\ket{rgrgrg\cdots}$ and $\ket{grgrgr\cdots}$) \cite{bluvstein2021controlling}.

In contrast, here we impose kinetic constraints using facilitation, which means that a Rydberg atom must be present to drive other nearby atoms to the Rydberg state \cite{ates2007antiblockade,marcuzzi2017facilitation,wintermantel2020unitary,magoni2021emergent}. Applying a global detuning facilitates Rydberg excitations, which lie on a shell of tunable radius given by $R_{f} = (C_{6}/\Delta)^{1/6}$ [Fig.~\ref{fig:schematic}(b)]. We use facilitation to selectively make beyond-nearest-neighbor interactions resonant. Combining facilitation with Rydberg blockade then allows us to achieve a broad class of locally constrained models.

\begin{figure*}[!htbp]
    \centering
    \includegraphics[keepaspectratio,width=18cm]{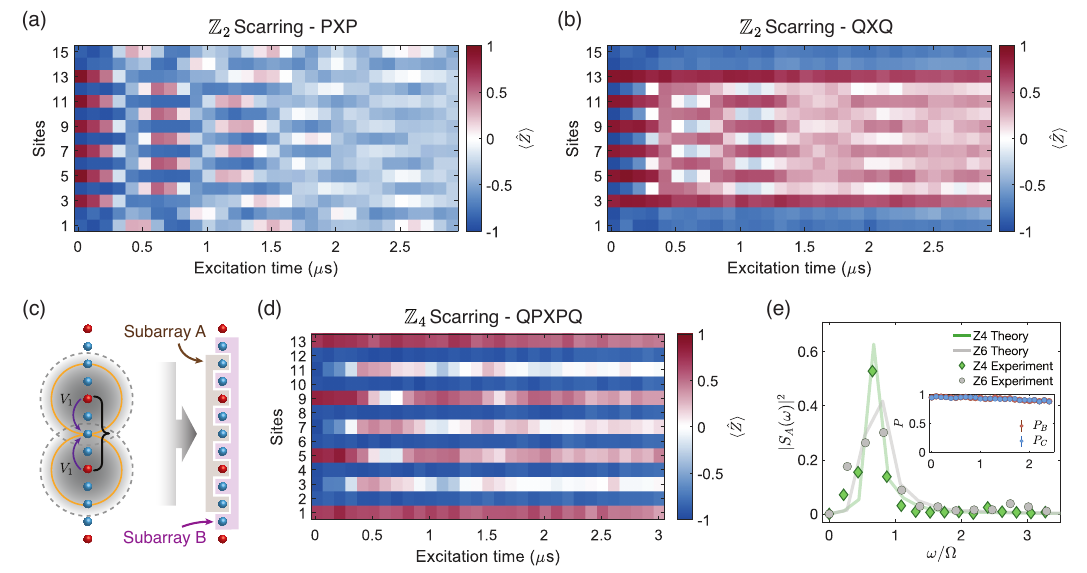}       \caption{Experimental observations of quantum many-body scarring in locally constrained models. Single-atom-resolved $Z$ measurements of spin chains in the (a) PXP and (b) QXQ models initialized in $\mathbb{Z}_{2}$ ordered states surrounded by ground state atoms. (c) The QPXPQ model is achieved by setting $\Delta = 2V_{1}$ with $V_{0}, V_{1} \gg \Omega$, such that an atom flips its spin only when flanked by nearest-neighbor ground state atoms and next-nearest-neighbor Rydberg excitations. Under the QPXPQ Hamiltonian, the dynamics of atoms in subarray $A$ (odd sites, boundary atoms excluded) are distinct from that in subarray $B$ (even sites). (d) Single-atom-resolved $Z$ measurements of the spin chain showing the $\mathbb{Z}_{4}$ quantum many-body scar at $\Delta = 2V_{1}$. (e) Fourier analysis of the $\mathbb{Z}_{2k}$ scars observed on subarray $A$. Experiment data (green diamonds for $k = 2$) show reasonable agreement with theory (green curve), which predicts a peak at $\omega \approx 0.67\Omega$. The $\mathbb{Z}_{6}$ scar manifests as a Fourier peak in subarray $A$ (gray circles for experimental data and gray curve for theory), alongside frozen ground state densities $P_{B/C}$ in the other two subarrays (inset). The theory calculations use $H_{\text{Ryd}}$ (Eq.~(\ref{eq:Hamiltonian})), where $\Omega=2\pi \times 1.45$~MHz, with $V_{1}=2\pi \times 5$~MHz for the $\mathbb{Z}_{4}$ case whereas $V_{2}= 2\pi \times 4.3$~MHz for the $\mathbb{Z}_{6}$ case.}
    \label{fig:scars}
\end{figure*}

To explain how Krylov subspaces emerge in our system, we first consider only facilitating nearest-neighbor interactions. Setting the global detuning to $\Delta = 2 V_{0}$, we now allow an atom to flip its state only when facilitated by two nearest neighbor Rydberg atoms. This constraint is captured by the effective Hamiltonian given by the QXQ model $H_{\text{QXQ}} = \frac{\Omega}{2}\sum_{i} Q_{i-1} X_{i} Q_{i+1}$.

At first glance, it appears that the QXQ model can be mapped exactly onto the PXP model by interchanging $|g\rangle \leftrightarrow |r\rangle$ \cite{valencia2023rydberg}. Indeed, we observe quantum many-body scars in both the QXQ and PXP models when a subset of the atom array is initialized in the $\mathbb{Z}_{2}$ ordered state  [Fig.~\ref{fig:scars}(a) and (b)]. However, while the PXP and QXQ models are symmetrically mapped to each other, the Hilbert space connectivity for experimentally accessible states is different for the two models. When the $\mathbb{Z}_{2}$ ordered subarray is surrounded by ground state atoms, the entire system takes part in the scarring dynamics under the PXP model. In contrast, for the QXQ model, the Rydberg atoms at the boundaries of the ordered state and atoms beyond the boundaries remain frozen, because these atoms do not fulfill the facilitation condition. In general, $\ket{gg}$ substrings, which can be easily prepared, would remain frozen within an atom chain under the QXQ model. In contrast, their symmetric mapping as frozen $\ket{rr}$ substrings in the PXP model would be experimentally challenging to prepare. As a result, the PXP subspace (also referred to as the primary $V_{0}$ block in Fig.~\ref{fig:schematic}(c)) is completely connected in the PXP model but fragmented in the QXQ model.

\section{Engineering Hilbert space fractures}

\begin{figure*}[!htbp]
    \centering
    \includegraphics[keepaspectratio,width=18cm]{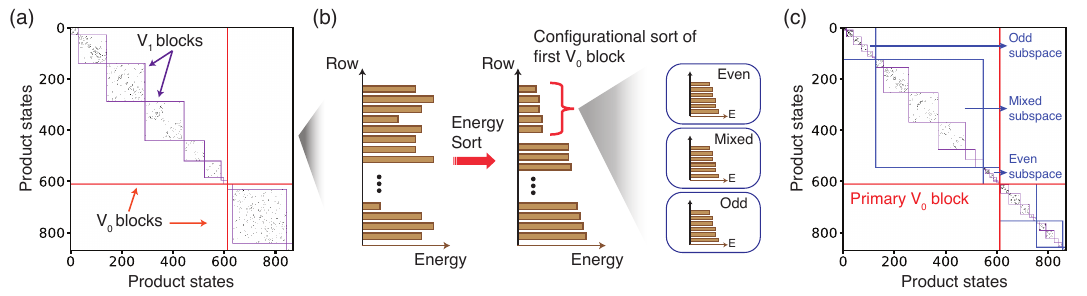}
    \caption{Hilbert space fragmentation of a 13-atom open chain. (a) A zoomed-in matrix plot of the QPXPQ Hamiltonian, with the product state basis sorted only by energy. (b) To uncover the Hilbert space fractures, we first sort the product-state basis in energy, from which we identify the first $V_{0}$ block. States in each $V_{0}$ block are then grouped according to their configuration, after which a final energy sort is performed within each configurational group. (c) The resultant matrix plot, focusing on the first $V_{0}$ energy block, after the product-state basis ordering. Disjointed blocks emerge due to different energy scales (red for $V_{0}$, purple for $V_{1}$) and configurational disconnectivity (blue).
    }
    \label{fig:matrixplots}
\end{figure*}

To further engineer the Hilbert space fractures, we now extend, both theoretically and experimentally, the range of kinetic constraints by using successive orders of the van der Waals interaction. Consider the regime where only the first two orders of the interaction are large ($V_0, V_1 \gg \Omega \gg V_2$). Setting $\Delta = 2 V_{1}$ then results in the model:
\begin{equation}
  H_{\text{QPXPQ}} = \frac{\Omega}{2}\sum_{i}Q_{i-2}P_{i-1}X_{i}P_{i+1}Q_{i+2}  
\end{equation}
Explicitly, this means that an atom will only flip its state when its immediate neighbors are ground state atoms and next-nearest neighbors are Rydberg atoms [Fig.~\ref{fig:scars}(c)]. This coexistence of both Rydberg blockade and facilitation constraints gives rise to qualitatively distinct dynamics in different sectors of our system, which we track by partitioning the bulk of the chain into two subarrays $A$ and $B$ to highlight their contrasting behavior [Fig.~\ref{fig:scars}(c)].

Under the QPXPQ model, we observe quantum many-body scars from $\mathbb{Z}_{4}$-ordered initial states, with the single-atom-resolved $Z$ measurements oscillating between two $\mathbb{Z}_{4}$ sectors of the chain, while atoms in the other $\mathbb{Z}_{2}$ sector (subarray $B$) remain completely frozen in the ground state [Fig.~\ref{fig:scars}(d)]. The scar oscillations manifest as a single peak in a Fourier analysis of the $\mathbb{Z}_{4}$ dynamics [Fig.~\ref{fig:scars}(e)]. 

We note that our approach enables us to access a broad class of locally constrained Hamiltonians $H_{\text{QPXPQ}}(k) = \frac{\Omega}{2}\sum_{i}Q_{i-k}X_{i}Q_{i+k}\prod_{j=1}^{k-1}P_{i-j}P_{i+j}$ by setting $\Delta = 2V_{k-1}$. Besides $k=2$ that corresponds to $\mathbb{Z}_{4}$ scarring, we also realize the case of $k=3$ by operating in the regime where $V_{0}$, $V_{1}$, and $V_{2}$ are all large. Here, the bulk of a 19-atom chain is partitioned into three interleaved subarrays ($A$, $B$, $C$) and initialized in a $\mathbb{Z}_{6}$ ordered state. In this case, we observe quantum many-body scars, as indicated by the Fourier peak in the dynamics between two $\mathbb{Z}_{6}$ sectors within subarray $A$, whereas the remaining subarrays remain frozen in the ground state [Fig.~\ref{fig:scars}(e)].

At this point, it is instructive to consider matrix plots of the QPXPQ Hamiltonian in the product state basis. Without loss of generality, we focus on the case of $k = 2$. The different interaction energy scales ($V_{0}$, $V_{1}$) provide a natural reference for ordering the basis so that basis states with the same energy can be grouped together as a single block. As such, it is natural to consider ordering the product-state basis of this chain by their energies with respect to $V_{0}$ and $V_{1}$ [Fig.~\ref{fig:matrixplots}(a)]. However, the dynamics in Fig.~\ref{fig:scars}(d) are highly dependent on the configuration of the initial state, which suggests the emergence of a Hilbert space structure unrelated to the energy windows.

To capture these observations, we take the product states in the first $V_{0}$ block and categorize them according to their configuration. We group product states in the even (odd) subspace if they have Rydberg excitations on only even (odd) sites, and the remaining product states fall in the mixed subspace. This configuration-based sorting gives rise to a secondary block structure as shown in Fig.~\ref{fig:matrixplots}(c). Within each configurational subspace, the Hilbert space is further split up according to the $V_{1}$ interaction energy scale. The fact that the odd subspace projects all atoms on even sites into the ground state and remains disconnected from other subspaces under $H_{\text{QPXPQ}}$ allows atoms in subarray $B$ stay frozen over time. 

While Fig.~\ref{fig:matrixplots} focuses on the experimentally accessible subspace (i.e.\ the first $V_{0}$ block), the number of fragmented sectors in the entire Hilbert space grows exponentially with the size of the chain. Briefly, the Hilbert space fragmentation can be understood in terms of how frozen segments separate a chain into multiple regions with independent dynamics, where these frozen segments can be formed by concatenating substrings comprising five spins that do not satisfy the projectors as ordered in the QPXPQ model (e.g.~$\ket{\cdots rgggg \cdots}$). Since each separated region evolves independently of other regions, there are a large number of disconnected subspaces (see also Appendix~\ref{Appendix:B.3}).

\begin{figure*}[!htbp]
    \centering
    \includegraphics[keepaspectratio,width=12.1cm]{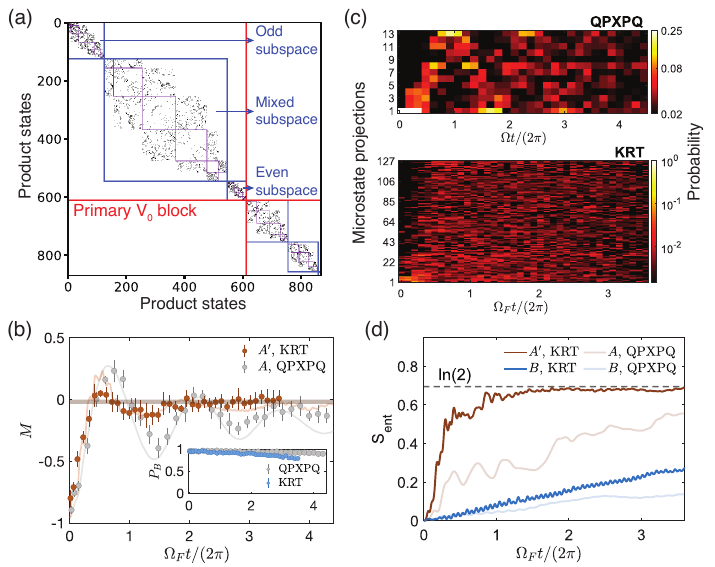}
        \caption{Signatures of Krylov-restricted thermalization. (a) Matrix plot under $H_{\text{KRT}}$ for a 13-atom linear chain. Relaxing the $V_{1}$ energy restriction allows thermalization to proceed within each configurational subspace while the three configurational subspaces remain disconnected. (b) Time-dependent behavior of the staggered magnetization for subarray $A$ or $A'$ and (inset) ground state density for subarray $B$, defined as $P_{B}(t) = \frac{1}{N_B}\sum_{i \in B} \langle \hat{P_i} \rangle$. Under $H_{\text{KRT}}$, $M_{A'}$ (brown data points with brown theory curve) quickly relaxes from -1 for an initial $\mathbb{Z}_{4}$ ordered state to its final thermal value near 0 (horizontal solid line). In contrast, under $H_{\text{QPXPQ}}$, $M_{A}$ displays persistent oscillations corresponding to a $\mathbb{Z}_{4}$ scar (gray data points with gray theory curve). In both cases, the atoms in subarray $B$ remain frozen in the ground state. (c) Microstate projections of experimentally measured bitstrings onto the constrained product state basis, ordered by their Hamming distance, under $H_{\text{QPXPQ}}$ and $H_{\text{KRT}}$. (d) Numerical plots of the averaged single-atom von Neumann entropy. For single atoms in subarray $A'$, $S_{\text{ent}}$ rapidly saturates to the thermal value for subarray $A'$ under $H_{\text{KRT}}$ (dark brown). Theory curves in (b) and (d) are calculated with $H_{\text{Ryd}}^{\text{FFM}}$ (Eq.~(\ref{eq:Ham_FFM})), where $\Omega = 2\pi \times 1.48$~MHz and $V_{1} = 2\pi \times 5$~MHz. 
        }
    \label{fig:KRT}
\end{figure*}

\begin{figure*}[!htbp]
    \centering
    \includegraphics[keepaspectratio,width=12.1cm]{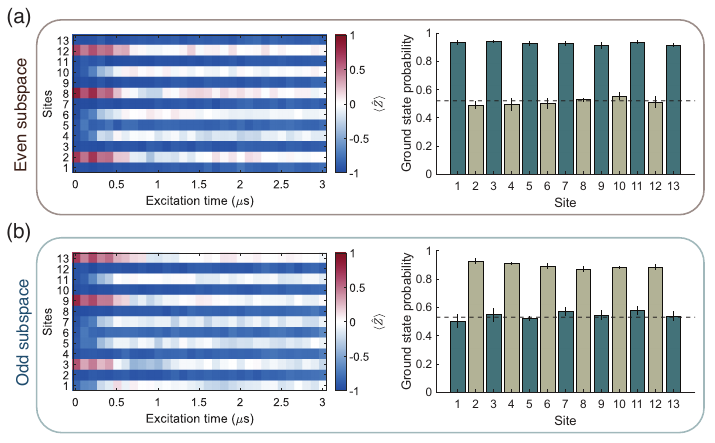}
        \caption{Thermalization within distinct subspaces. (a) Thermalization within the even subspace. Single-atom-resolved dynamics (left) for an array initialized with Rydberg excitations on sites (2, 8, 12). The site-resolved ground state probabilities, time-averaged from 2.7~$\mu$s to 3~$\mu$s (corresponding to $\Omega_{F} t/(2\pi) = 1.7 - 1.9$, where $S_{\text{ent}}$ has already saturated), are shown on the right. Dark (light) green bars indicate the odd (even) subarray. The horizontal dashed line indicates the thermal predicted value. (b) Corresponding dynamics and long-time-averaged ground state probabilities, shown to depict thermalization within the odd subspace. Here the array is initialized with Rydberg excitations on sites (3, 9, 13). Despite the two initial states of (a) and (b) having the same energy, they do not thermalize with each other.}
    \label{fig:subspace}
\end{figure*}

The disconnectivity between these configurational subspaces is no longer present for the blockade-only PPXPP Hamiltonian (Appendix~\ref{Appendix:B.5}), highlighting that facilitation is a necessary ingredient for realizing clean Hilbert space fractures in the primary $V_{0}$ block. Intuitively, this arises from the fact that facilitation allows one to selectively excite atoms only at a given distance, whereas Rydberg blockade acts on all atoms within a sphere and therefore lacks such distance selectivity. In other words, Krylov subspaces in our system arise from strong kinetic constraints in the form of conditional excitations from both facilitation and blockade that invoke higher-order interaction terms. Such a fragmented system is distinct from existing proposed models.

\section{Krylov-restricted thermalization}
Importantly, Fig.~\ref{fig:KRT}(a) predicts that the configurational subspaces remain disconnected upon coupling states that are previously separated in energy by $V_{1}$. We effect the coupling by adding, on top of the first drive at $\Delta = 2V_{1}$, a second drive detuned at $\Delta' = V_{1}$. The two-drive experiments are implemented using Floquet frequency modulation \cite{zhao2023floquet}, realizing the effective Hamiltonian (see Appendix~\ref{Appendix:B.2}):
\begin{equation}
\label{eq:KRT_main}
    \begin{split}
        H_{\text{KRT}} &= \frac{1}{2}\sum_{j}P_{j-1}X_{j}P_{j+1}\left(-\Omega_{F} Q_{j-2}Q_{j+2} \right. \\
        &\quad \left. + i\Omega_{F}' (P_{j-2}Q_{j+2} + Q_{j-2}P_{j+2})\right)
    \end{split}
\end{equation}
Here, $\Omega_{F}$ and $\Omega_{F}'$ are the Rabi frequencies corresponding to the drives at $\Delta = 2V_{1}$ and $\Delta' = V_{1}$, respectively.

To contrast the dynamics under $H_{\text{KRT}}$ versus under $H_{\text{QPXPQ}}$, we initialize a 13-atom chain in the $\mathbb{Z}_{4}$ ordered state, where the atoms are partitioned in the same way as depicted in Fig.~\ref{fig:scars}(c). Strikingly, instead of observing generic thermalization across the entire array, we find that the atoms in subarray $A$ quickly thermalize, whereas atoms in subarray $B$ remain frozen in the ground state. In fact, introducing the additional $\Delta' = V_{1}$ drive couples the boundary atoms to subarray $A$ and allows them to take part in thermalization, unlike in the QPXPQ framework where boundary Rydberg atoms remain frozen and do not take part in scar dynamics. We refer to the collection of atoms in subarray $A$, together with the two boundary atoms, as subarray $A'$. The staggered magnetization for this boundary-included subarray is defined as:
\begin{equation}
    M_{A'}(t) = \frac{1}{N_{A'}} \sum_{i\in A'} (-1)^{i} \langle Z_{i}(t) \rangle \, ,
\end{equation}
where $i$ indexes the atom on odd sites of the entire chain. (The corresponding staggered magnetization definition for subarray $A$ under the QPXPQ Hamiltonian is given in Appendix~\ref{Appendix:C.2}.) The restricted-thermalization dynamics can be perceived through a few different measures: comparing the time-dependent staggered magnetization of subarray $A'$ versus the ground state density for subarray $B$, tracking the trajectories of experimentally observed microstate projections, computing the von Neumann entropy of single atoms in the two subarrays.

Specifically, under $H_{\text{KRT}}$, Fig.~\ref{fig:KRT}(b) shows that $M_{A'}$ quickly decays to the thermal value as predicted by a microcanonical ensemble (see Appendix~\ref{Appendix:C.1}), whereas the ground state density for subarray $B$ remains high over time. Further, the microstate projections are observed to have a near-uniform spread across the basis of $(2^{N_{A'}} - 1)$ states [Fig.~\ref{fig:KRT}(c)]. This is in stark contrast to the $\mathbb{Z}_{4}$ scar behavior, which features persistent revivals in the staggered magnetization $M_{A}$ and a periodic trajectory in the constrained product-state basis for the QPXPQ model [Fig.~\ref{fig:KRT}(c)]. Finally, the averaged single-atom von Neumann entropy $S_{\text{ent}}$ for atoms in subarray $A'$ is found to quickly saturate to the thermal value of $\ln 2$, whereas it grows much more slowly for atoms in subarray $B$. The finite slow decay of $P_{B}$ and correspondingly slow growth of $S_{\text{ent}}^{B}$ for subarray $B$ is attributed to deviations from the effective model Hamiltonians (Appendix~\ref{Appendix:D.1}).  

We note that our findings are distinct from previous studies of thermalization with Rydberg atom arrays \cite{olmos2010thermalization,ates2012thermalization,kim2018detailed,bluvstein2021controlling,choi2023preparing}, where states are found to be entangled with all other states within the same energy window \cite{bluvstein2021controlling,choi2023preparing}. Our observations of strongly site-dependent thermalization, which occurs in the absence of any site-selective addressing, is a manifestation of the configurational disconnectivity due to Hilbert space fragmentation (Appendix~\ref{Appendix:B.4}). 

Notably, states in the odd configurational subspace never thermalize with any states in the even or mixed configurational subspaces, and vice versa, even when they are \textit{exactly} degenerate in energy [Figs.~\ref{fig:KRT}(a),~\ref{fig:subspace}]. This is the smoking-gun signature of Krylov-restricted thermalization, which constitutes a clear violation of the ETH for the full Hilbert space at this energy window, even though we see signatures of thermalization within a given subspace.

\section{Conclusion}

We have demonstrated the first realization of Krylov-restricted thermalization, by which we experimentally resolve the longstanding tension between thermalization and memory. Our results prompt a relook at the cornerstone ideas of quantum thermalization \cite{deutsch1991quantum,rigol2008thermalization, neill2016ergodic,kaufman2016quantum}, which may have more stringent requirements beyond energy considerations. Our work can be extended to yield a comprehensive description of Krylov subspaces, which would be important for elucidating the full nature of ETH violation \cite{rakovszky2020statistical,moudgalya2021thermalization,moudgalya2022hilbert}. Further, while kinetic constraints give rise to both $\mathbb{Z}_{2k}$ scars and Hilbert space fragmentation in our models, the direct relationship between scars and fragmentation may be complex and should be systematically explored in future work.

Our observations open the door to comparisons against other fragmented systems such as dipole-conserving models \cite{sala2020ergodicity,khemani2020localization,kohlert2023exploring}, where Krylov-restricted thermalization can give rise to intriguing fracton dynamics \cite{feng2022hilbert,adler2024observation,boesl2311deconfinement}. On the other hand, explorations of non-ergodic dynamics in fragmented systems would provide valuable insights on dynamical phases such as constraint-driven many-body localization (MBL) \cite{chen2018how,tomasi2019dynamics,herviou2021many} and non-Hermitian skin clusters \cite{shen2022non}. 

The study of restricted dynamics in fragmented systems may find interesting applications beyond quantum simulation. For instance, Krylov-restricted dynamics can potentially motivate the construction of quantum error codes where errors remain confined to specific subspaces. Further, variational algorithms or quantum optimization algorithms may leverage on the constrained evolution in Krylov subspaces to reduce the computational complexity \cite{mcclean2017hybrid,yoshioka2022generalized,motta2020determining, colless2018computation,mcclean2020decoding}. Krylov-restricted thermalization could also lead to the development of quantum sensors with configuration-sensitive detection, offering efficient readout over a reduced basis \cite{li2023improving}. More generally, the control of information dynamics in fragmented systems can be expanded through out-of-time-ordered correlators \cite{lewis2019unifying,hahn2021information} and may be used to enhance quantum information processing and quantum metrology.

\begin{acknowledgments}
We thank Adam Kaufman, Wen Wei Ho, Ching Hua Lee, and Travis Nicholson for helpful discussions, as well as Fan Jia, Wen Jun Wee, An Qu, and Jiacheng You for technical assistance with the experiment setup. This research is supported by the National Research Foundation, Singapore and A*STAR under its Quantum Engineering Programme (NRF2021-QEP2-02-P09) and its CQT bridging grant.
\\
\paragraph*{Note added.}---During the completion of this work, we became aware of related theory work \cite{yang2024probing}.
\end{acknowledgments}
\appendix

\section{Experiment methods}
\subsection{Setup and sequence}
\label{Appendix:A.1}

We first prepare a two-dimensional array \cite{tian2023parallel} of 57--75 static tweezer traps by sending 808~nm light through a pair of acousto-optic deflectors (AODs) and focusing the output beams to a waist of 0.98(2)~$\mu$m through a microscope objective of numerical aperture NA = 0.5. Through light shift measurements, we determine the average trap depth to be 1.0~mK with 2-5\% inhomogeneity (relative standard deviation) across the array. Using $D_{1}$ $\Lambda$-enhanced loading, we load the static array of traps with single $^{87}$Rb atoms with $\ge$80\% loading probability. Single-atom detection is performed by fluorescence imaging using optical molasses beams with 99.39(3)\% fidelity. We employ a multitweezer algorithm utilizing an array of 852~nm mobile tweezers to assemble three identical defect-free atom chains. The defect-free success probability is about 80\% for each chain. In all of our experiments, these chains are spaced about 3$R_{b}$ apart such that interactions between the chains are on the scale of kHz and can be considered negligible compared to the Rabi frequency, which is on the scale of MHz.

To reduce the in-trap atomic temperature, we perform two stages of cooling. We first employ $D_{1}$ $\Lambda$-enhanced gray molasses to cool the atoms down to 37(2)~$\mu$K, followed by a preliminary implementation of Raman sideband cooling in 0.6~mK traps for 8~ms. The final atom temperature is 19(1)~$\mu$K as determined from comparing the measured release-and-recapture probability against Monte Carlo simulations. After cooling, we initialize the atoms in the ground state $\ket{g} = \ket{5 S_{1/2}, F = 2, m_{F} = 2}$ with 99\% fidelity by optically pumping the atoms in the presence of a quantization magnetic field of 3.7~G. The tweezer traps are subsequently turned off during Rydberg excitation.

Our Rydberg excitation follows a two-photon scheme, where a global pulse consisting of a $\sigma^{+}$-polarized 420~nm beam and a counter-propagating $\sigma^{-}$-polarized 1013~nm beam is sent along the quantization axis to excite the atoms to the Rydberg state. The 420~nm laser is red-detuned from the $\ket{6 P_{3/2}, F = 3, m_{F} = 3}$ intermediate state by $\Delta' \approx 2\pi \times 780$~MHz, yielding a combined two-photon Rabi frequency of around $\Omega = 2\pi \times 1.4$~MHz. The Rydberg beam alignment is kept stabilized by piezoelectric mirrors. The damping time for the ground-Rydberg Rabi oscillations of a single atom is measured to be $\tau = 27(6)~\mu$s and is well beyond the time scale of dynamics studied in this work, indicating that single-atom decoherence \cite{leseleuc2018analysis,tamura2020analysis} is not a limitation in our study.

For most of the experiments, we excite the atoms to the $\ket{70 S_{1/2}, m_{J} = 1/2}$ Rydberg state. For the QXQ experiment, the interatomic spacing is set to $a = 7.46~\mu$m, resulting in a measured interaction strength of $V_{0} = 2\pi\times4.9(1)$~MHz. This spacing is changed to 3.73~$\mu$m for the QPXPQ and Krylov-restricted thermalization experiments. For the $\mathbb{Z}_{6}$ scarring experiment, we instead couple the atoms to the $\ket{87 S_{1/2}, m_{J} = 1/2}$ Rydberg state. In this case, we use only one defect-free 19-atom chain (as opposed to three defect-free chains) with $a = 3.89~\mu$m to obtain $V_{2} = 2\pi \times 4.3(1)$~MHz. 

We implement single-site addressing with the same 852~nm laser used for atom rearrangement. The single-site addressing beam is applied alongside a global Rydberg $\pi$-pulse for initial state preparation. The dynamics of interest are probed with a second global pulse of varying duration, after which atoms in the ground state are recaptured into 1~mK tweezer traps while an additional 3~$\mu$s microwave pulse is applied to ionize the Rydberg atoms. The imaging beams are applied a final time to read out the ground (Rydberg) state as the presence (absence) of an atom. We realize a Rydberg detection fidelity of $97(2)\%$. Each data point shown in this manuscript is averaged from at least 200 defect-free chain samples.

For future experiments, larger array sizes are desired to test quantum many-body theory in a regime where full numerical simulations are challenging to perform. We note that current array sizes are primarily limited by the finite state preparation and detection efficiency. A secondary limit on the array size comes from the finite bandwidth of the acousto-optic deflectors. The latter can be circumvented by reducing the interaction scale with a smaller principal quantum number for the Rydberg state or by increasing the Rabi frequency.

\subsection{Floquet frequency modulation}
\label{Appendix:A.2}

In our experiments, we use Floquet frequency modulation (FFM) to generate the two drives necessary to realize $H_{\text{KRT}}$. Modulating the detuning of a single drive at frequency $\omega_d$ as $\Delta(t) = \Delta_0\sin{\omega_dt}$ yields the Hamiltonian
\begin{eqnarray}
\label{eq:Ham_FFM}
H_{\text{Ryd}}^{\text{FFM}} &=& -(\Delta_0\sin{\omega_d t}) \sum_{i=1}^{N} Q_{i} + \frac{\Omega}{2}\sum_{i=1}^{N} X_{i} \nonumber \\
    && + \sum_{j = 1}^{N-1} \sum_{i = 1}^{N-j} V_{j-1} Q_{i} Q_{i+j} \, .
\end{eqnarray}
In a rotating frame, the Rabi frequency is modified by the frequency modulation: 
\begin{equation}\label{FFM}
    \Omega \xrightarrow{\text{FFM}} \Omega \sum_{m=-\infty}^{\infty}i^mJ_m(\alpha)e^{im\omega_dt} \, ,
\end{equation}
where $\alpha = \Delta_0/\omega_d$ is the modulation depth. To realize $H_{\text{KRT}}$, we set $\omega_d = V_{k-1}$ and $\alpha = 2.4$, the first zero of $J_0$. This ensures that the carrier $m=0$ component vanishes, while keeping similar amplitudes $J_1(\alpha) = 0.52$ and $J_2(\alpha) = 0.43$ for the components in Eq.~(\ref{FFM}) that are detuned at $\Delta' = V_{k-1}$ and $\Delta = 2V_{k-1}$, respectively. In other words, the Rabi frequencies for the two drives in Eq.~(\ref{eq:KRT_main}) are $\Omega_{F}' = 0.52\Omega$ and $\Omega_{F} = 0.43\Omega$, respectively. 

In practice, we implement Floquet frequency modulation by sending a focused 420 nm Rydberg laser (beam waist 25~$\mu$m) through an acousto-optic modulator (AOM) in a double-pass configuration. The AOM is in turn driven by an arbitrary waveform generator to achieve a time-varying frequency $\Delta(t)$. We calibrate the modulation by monitoring an optical beat note between the first-order deflected beam with a reference beam and find that the modulation depth deviates from the expected value by a linear scaling factor $\alpha/\alpha_{\text{actual}} = 1.01$. In addition, we minimize the residual amplitude modulation by compensating for the AOM frequency-dependent diffraction efficiency and by applying a waveform with an optimal set of amplitudes corresponding to the first four harmonics of the modulation frequency \cite{zhao2023floquet}.

We note that in principle, the Floquet frequency modulation can alter the off-resonant scattering rate from the intermediate state in the two-photon Rydberg excitation by 0.6\%. However, the non-modulated off-resonant scattering rate already exerts a negligible effect on the dynamics. Further, the atoms sample the same spatial profile of Rydberg laser intensities regardless of whether the modulation is effected. Therefore we do not expect an increase in decoherence when modulation is introduced.

\section{Effective models and discussion}
\subsection{Symmetry between QXQ and PXP}
\label{Appendix:B.1}

For a 1D chain of equally spaced atoms, the nearest-neighbor approximation to the Rydberg Hamiltonian is given by $(\hbar = 1)$:
\begin{equation} \label{H_NN}
    H = -\Delta \sum_{i=1}^{N} Q_{i} + \frac{\Omega}{2}\sum_{i=1}^{N} X_i + \sum_{i=1}^{N-1} V_{0} Q_{i} Q_{i+1}
\end{equation}
Setting $\Delta = 2V_0$ and making the substitution $Q_{i} = \mathbb{I}-P_{i}$, the Hamiltonian transforms to (neglecting constant energy offsets):
\begin{equation} \label{H_sym}
    H = \frac{\Omega}{2}\sum_{i=1}^{N} X_{i} + \sum_{i=1}^{N-1} V_{0} P_{i}P_{i+1} - V_0 (Q_{1}+Q_{N})
\end{equation}
The last term is a result of our chain having open boundaries, such that atoms at the ends interact with only one instead of two other atoms. The similar forms of equations~(\ref{H_NN}) and (\ref{H_sym}) reveal the symmetry between the QXQ and PXP models in the bulk. 

\subsection{Effective Hamiltonians with beyond-nearest-neighbor interactions}
\label{Appendix:B.2}
We now consider beyond-nearest-neighbor interactions and derive effective Hamiltonians for our systems in the regime where $V_0 , V_1 ... V_{k-1} \gg \Omega \gg V_{k}$. In other words, we consider the scenario where the first $k$ orders of interaction are significant to the dynamics at hand ($k = 1$ corresponds to the nearest-neighbor case). To capture the essential physics, we work with periodic boundary conditions. In this regime, the Hamiltonian is:
\begin{equation} \label{H_generic}
    H = -\Delta \sum_{i=1}^{N} Q_{i} + \frac{\Omega}{2}\sum_{i=1}^{N} X_{i} + \sum_{j=1}^{k}\sum_{i=1}^{N} V_{j-1} Q_{i} Q_{i+j} \, .
\end{equation}
We then enter the rotating frame with respect to the van der Waals interaction and detuning terms. This corresponds to a unitary transform $U = e^{it\sum_{j=1}^{k}\sum_{i=1}^{N} V_{j-1}  Q_{i} Q_{i+j}-it\Delta\sum_{i=1}^{N} Q_{i}}$. The rotated Hamiltonian is:
\begin{equation}
    H' = \frac{\Omega}{2}\sum_{i=1}^{N} e^{it[-\Delta+\sum_{j=1}^{k} V_{j-1} (Q_{i+j}+Q_{i-j})]}\sigma_{+}^i + \text{H.c.}
\end{equation}
where $\sigma_{+}^{i} =\ket{r_{i}}\bra{g_{i}}$ is the Pauli spin raising operator. Standard commutation relations between Pauli matrices and the Baker-Hausdorff lemma are used to arrive at the equality. Further, using the Pauli identity $e^{ia\vec{n}\cdot \vec{\sigma}} = \mathbb{I}\cos{a}+ i\vec{n}\cdot\vec{\sigma}\sin{a}$ and making the substitution $Q_{i} = \mathbb{I} - P_{i}$ where appropriate, we arrive at the expression:
\begin{widetext}
\begin{equation}\label{general_case}
    H' = \frac{\Omega e^{-i\Delta t}}{2}\sum_{i=1}^{N}\sigma^{i}_{+}\prod_{j=1}^k\left[\left(P_{i-j}P_{i+j}\right) + e^{iV_{j-1}t}\left(Q_{i-j}P_{i+j}+P_{i-j}Q_{i+j}\right) + e^{2iV_{j-1}t}\left(Q_{i-j}Q_{i+j}\right)\right] + \text{H.c.} 
\end{equation}
In the case of $k=1$ (nearest-neighbor interaction only), the above general expression simplifies to:
\begin{equation}
    H' = \frac{\Omega e^{-i\Delta t}}{2}\sum_{i=1}^{N}\sigma^{i}_{+}\left[\left(P_{i-1}P_{i+1}\right) + e^{iV_{0}t}\left(Q_{i-1}P_{i+1}+P_{i-1}Q_{i+1}\right) + e^{2iV_{0}t}\left(Q_{i-1}Q_{i+1}\right)\right] + \text{H.c.} 
\label{eqn:simplified}
\end{equation}

For the remainder of this section (equations~(\ref{pxp})--(\ref{H_KRT})), we apply a rotating-wave approximation while ignoring far off-resonant terms. This is valid in the case where $V_{0}, \ldots , V_{k-1} \gg \Omega$. For $k = 1$, we get the PXP model under $\Delta = 0$:
\begin{equation}\label{pxp}
    H_{\text{PXP}} = \frac{\Omega}{2}\sum_{i=1}^{N}P_{i-1} X_{i} P_{i+1} \, ,
\end{equation}
whereas setting $\Delta = 2V_0$ gives us the QXQ model:
\begin{equation}\label{qxq}
    H_{\text{QXQ}} = \frac{\Omega}{2}\sum_{i=1}^{N}Q_{i-1} X_{i} Q_{i+1} \, .
\end{equation}
In the general case of $k\geq 2$, setting $\Delta = 2V_{k-1}$ yields the generalized QPXPQ model as referenced in the main text:
\begin{equation}
    H_{\text{QPXPQ}}(k) = \frac{\Omega}{2}\sum_{i=1}^{N}Q_{i-k}X_{i}Q_{i+k}\prod_{j=1}^{k-1}P_{i-j}P_{i+j} \, .
\end{equation}
Starting from a state with $\mathbb{Z}_{2k}$ ordering, under this Hamiltonian, we would then expect only the atoms in the same $\mathbb{Z}_k$ sector to undergo dynamics, while the rest of the atoms should remain frozen in the ground state. In this constrained Hilbert space, the model may be mapped to a QXQ model on a smaller $m$-atom chain, which explains the scarring behavior that is observed. For our experiments with an $N$-atom open chain [Figs.~\ref{fig:scars}(c) and (d)], $m = (N + k - 1)/k$.

From Eq.~(\ref{general_case}), our motivation for adding a second drive detuned at $\Delta = V_{k-1}$ becomes clear: this drive is resonant for atoms that are blockaded by exactly one Rydberg atom $k$ sites away. Added to the original drive detuned at $\Delta = 2V_{k-1}$, this second drive loosens the kinetic constraints exactly $k$ sites away from the atom, while still maintaining blockade restrictions less than $k$ sites away. Accounting for the sideband phases from the Floquet frequency modulation drive, the effective Hamiltonian becomes:
\begin{equation} \label{H_KRT}
    H_{\text{KRT}}(k) = \frac{1}{2}\sum_{j=1}^{N}\left(-\Omega_{F} Q_{j-k}Q_{j+k}+i\Omega_{F}' (Q_{j-k}P_{j+k} + P_{j-k}Q_{j+k})\right)X_{j}\prod_{\ell=1}^{k-1}P_{j-\ell}P_{j+\ell} \, ,
\end{equation}
where $\Omega_{F}$ and $\Omega_{F}'$ are the Rabi frequencies corresponding to the drives detuned at $2V_{k-1}$ and $V_{k-1}$, respectively. In the main text, we assume $k=2$ wherever the argument $k$ is not explicitly included.

For a linear chain with open boundary conditions (OBC), the effective QPXPQ model (k=2) becomes:
\begin{equation} \label{eq:H_QPXPQ_OBC}
    H_{\text{QPXPQ}}^{\text{(OBC)}} = \frac{\Omega}{2}\sum_{j=3}^{N-2}Q_{j-2}P_{j-1}X_{j}P_{j+1}Q_{j+2} \, ,
\end{equation}
whereas the effective two-drive Krylov-restricted thermalization Hamiltonian becomes:
\begin{equation} \label{eq:H_KRT_OBC}
    \begin{aligned}
        H_{\text{KRT}}^{\text{(OBC)}} = \frac{1}{2} \bigg[ & \sum_{j=3}^{N-2}P_{j-1}X_{j}P_{j+1}\left(-\Omega_{F} Q_{j-2}Q_{j+2} + i\Omega_{F}' (Q_{j-2}P_{j+2} + P_{j-2}Q_{j+2})\right) \\
        & + i\Omega_{F}' \left(X_{1}P_{2}Q_{3} + P_{1}X_{2}P_{3}Q_{4} + Q_{N-3}P_{N-2}X_{N-1}P_{N} + Q_{N-2}P_{N-1}X_{N}\right) \bigg] \, .
    \end{aligned}
\end{equation}
\end{widetext}

\subsection{Hilbert space fragmentation under $H_{\text{QPXPQ}}$}
\label{Appendix:B.3}

In this section, we discuss the QPXPQ model and its unusual nonergodic dynamics, which we identify as a consequence of extensive fragmentation of the Hilbert space into many disconnected subspaces. Referencing previous works \cite{sala2020ergodicity}, we briefly examine the resulting frozen states and the spatially disconnected regions separated by these frozen components.

There are exponentially many product states in the local $Z$-basis that are invariant under the time evolution operator, referred to as frozen states. The simplest case is the polarized state $\ket{0} \equiv \ket{\cdots ggggg \cdots}$, remaining frozen due to the $Q$ projectors never being satisfied. Similarly, $\ket{R} \equiv \ket{\cdots rrrrr \cdots}$ remains frozen due to the $P$ projectors. Based on the ordering of projectors, a list of frozen substrings consisting of five spins can be constructed: $\ket{\cdots rgggg \cdots}$, $\ket{\cdots ggggr \cdots}$, $\ket{\cdots rggrg \cdots}$, $\ket{\cdots grggr \cdots}$, $\ket{\cdots grrrg \cdots}$, $\ket{\cdots rrggr \cdots}$, $\ket{\cdots rggrr \cdots}$. By adding frozen substrings on top of the polarized state, another frozen state is obtained. From the construction we deduce that the quantity of these states grows exponentially with the size of the system. We have demonstrated that substrings of certain patterns remain invariant under the action of $H_{\text{QPXPQ}}$. If we now consider a region between two such frozen substrings, this region stays disconnected from the rest of the chain. One can then cover the entire chain with many such regions of varying sizes and locations, surrounded by any such frozen substrings. Each of these regions has its own independent dynamics, giving rise to a large number of disconnected subspaces.

\subsection{Hilbert space fragmentation under $H_{\text{KRT}}$}
\label{Appendix:B.4}
\begin{figure}[t]
    \centering
    \includegraphics[keepaspectratio,width=8.8cm]{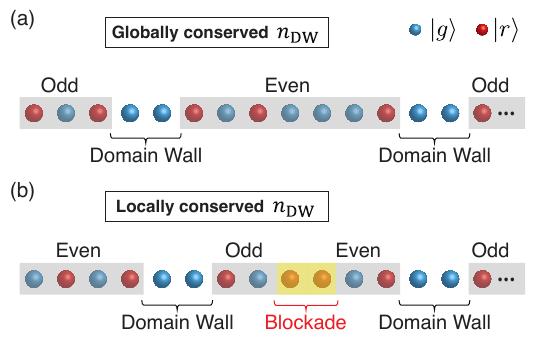}
    \caption{Domain walls under $H_{\text{KRT}}$. (a) Within the primary $V_{0}$ block, one can partition the chain into subchains of alternating parity. The number of domain walls $n_{\text{DW}}$ between partitions is globally conserved. $n_{\text{DW}}$, along with the parity of the leftmost subchain, uniquely identifies a subspace. (b) Outside the primary $V_{0}$ block, $n_{\text{DW}}$ is no longer globally conserved, but is instead locally conserved for each sector of the chain separated by $\ket{rr}$ blocks.
    }
    \label{edfig:krtdw}
\end{figure}
Central to our observation of Krylov-restricted thermalization is the Hilbert space fragmentation exhibited under $H_{\text{KRT}}$. Fragmented systems cannot be characterized by conventional symmetries \cite{moudgalya2021thermalization,moudgalya2022quantum,moudgalya2022hilbert}. Hence, a natural question to ask is the following: is the system truly fragmented, or are there in fact conventional symmetries that uniquely identify each Krylov subspace? 

To address this question, we first confine ourselves to the primary $V_{0}$ block. We observe that every $Z$-basis product state comprises subchains separated by domain walls, which we identify as clusters of even-numbered ground-state atoms [Fig.~\ref{edfig:krtdw}(a)]. Adjacent subchains have alternating parity, which denote whether Rydberg atoms reside on only odd or even sites of the entire chain. (In the main text, the ``odd'' or ``even'' subspace refers to the subspace of product states with no domain walls.)

Since spin-flip processes are allowed only when the nearest Rydberg atom is two sites away, we heuristically see that the number of these domain walls, $n_{\text{DW}}$, remains conserved under $H_{\text{KRT}}$. This observation prompts us to construct an operator that counts these domain walls, given by the following expression for periodic boundaries:
\begin{widetext}
    \begin{equation} \label{eq:ndw}
        n_{\text{DW}} = \frac{n_p}{2} + \frac{1}{2}\sum^{s}_{i=1} \Big[ \left|Q_{2i-1}P_{2i} - P_{2i}Q_{2i+1}\right| +\left|Q_{2i}P_{2i+1} - P_{2i-1}Q_{2i}\right| \Big] \, ,
    \end{equation}
\end{widetext}
where $n_p = 0$ and $s = N/2$ for $N$ even and $n_p = Q_{N}P_{1} - P_{N}Q_{1}$ and $s = (N-1)/2$ for $N$ odd. Intuitively, one of the four terms in the square brackets of Eq.~(\ref{eq:ndw}) yields a contribution equal to unity whenever one encounters or leaves a domain wall. On the other hand, for any subchain with Rydberg atoms on only the odd (even) sites, the first and second (third and fourth) terms contribute equally and cancel each other out. Thus, for a given $n_{\text{DW}}$, the subspace splits into exactly two, each characterized by the parity of the leftmost subchain.

At this point, we have seemingly characterized the subspaces with a global quantum number, which would have implied that the model is not fragmented! However, the above description fails when considering states outside the primary $V_{0}$ block. The presence of $\ket{rr}$ blocks prevents any entanglement from spreading between sections of the chain and $n_{\text{DW}}$ generally fails to uniquely characterize even a finite number of subspaces. The number of domain walls is now only conserved locally for individual sectors of the chain separated by $\ket{rr}$ blocks [Fig.~\ref{edfig:krtdw}(b)]. Since there are already exponentially many ways to uniquely blockade the chain into subchains \cite{moudgalya2022quantum}, the Krylov-restricted thermalization model is indeed fragmented.

\subsection{PPXPP model versus QPXPQ model}
\label{Appendix:B.5}

\begin{figure}[t]
    \centering
    \includegraphics[keepaspectratio,width=8.8cm]{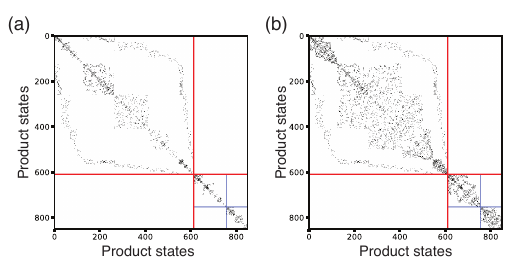}
    \caption{PPXPP model. (a) A zoomed-in matrix plot with the boundaries of the first $V_{0}$ block marked out. Basis sorting is performed as per Fig.~\ref{fig:matrixplots}. (b) The corresponding matrix when a second drive at $\Delta = V_1$ is incorporated. We do not see the emergence of any configurational subspaces in the first $V_{0}$ block.
    }
    \label{edfig:ppxpp}
\end{figure}

In the main text, we noted that facilitation is a necessary ingredient to observe the emergence of configurational subspaces in the first $V_{0}$ block. To demonstrate this, we consider the converse situation: the blockade-only PPXPP model and its corresponding effective Hamiltonian when a second drive at $\Delta = V_1$ is added, following the same basis ordering procedure as described in Fig.~\ref{fig:matrixplots}. As expected, after ordering the product states only based on their energies, a block structure emerges corresponding to the $V_0$ and $V_1$ energy windows. However, a configuration sort as performed in Fig.~\ref{fig:matrixplots} does not result in clear disconnected blocks remaining in the first $V_{0}$ block after a second drive is added with detuning $\Delta = V_{1}$ [Fig.~\ref{edfig:ppxpp}(a) and (b)], in contrast to the case of $H_{\text{KRT}}$. This explains the drop in Fig.~\ref{edfig:model_dev}(a).
We note that a block structure does emerge in the second $V_0$ block. In this case, the configurational ordering arises purely from $ |rr\rangle$ sectors that remain frozen, which is a generic feature of both the PPXPP and QPXPQ models. On the other hand, states in the second $V_{0}$ block can be challenging to access experimentally, so in this work we have focused our efforts on the first $V_{0}$ block. 

\section{Krylov-restricted thermalization}
\subsection{Predictions from a microcanonical ensemble}
\label{Appendix:C.1}

\begin{figure}[t]
    \centering
    \includegraphics[keepaspectratio,width=8.8cm]{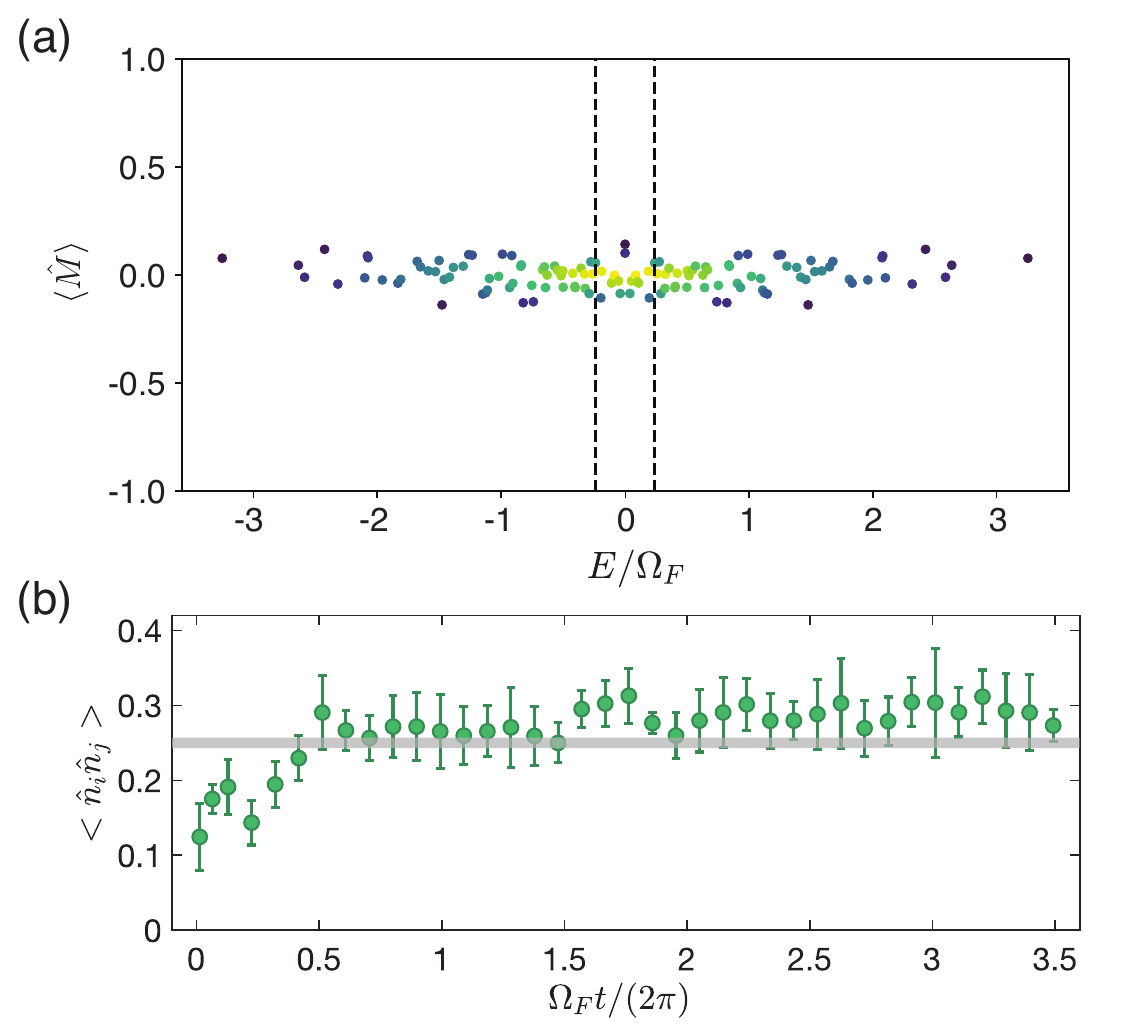}
    \caption{Thermalization in subspace. (a) Distribution of $\langle \hat{M} \rangle$ for each eigenstate, plotted as a function of the eigenenergy of $H_{A'}$. The two dashed lines mark the energy window $[-0.24\Omega_{F},0.24\Omega_{F}]$ used to obtain the microcanonical ensemble. (b) Relaxation dynamics of the local two-body observable $\langle\hat{n}_{i}\hat{n}_{j}\rangle$. The horizontal line indicates the expected value for a thermal state.}
    \label{edfig:veri}
\end{figure}

To predict the thermal values shown in Figs.~\ref{fig:KRT}(b) and \ref{fig:subspace}, we compute the statistical average of the quantity of interest (e.g.\ the staggered magnetization expectation value or the ground state probability) for a microcanonical ensemble. For illustration purposes, we focus on the 13-atom chain thermalizing in the odd configurational subspace, such that the Hamiltonian for the seven atoms on the odd sites (constituting subarray $A'$) is given by:
\begin{widetext}
    \begin{equation} \label{H_eff_odd}
        H_{A'} = \frac{1}{2}(i\Omega_{F}' X_{1}Q_{2}+i\Omega_{F}' Q_{6}X_{7}+\sum_{j=2}^{6}\left(-\Omega_{F} Q_{j-1}Q_{j+1}+i\Omega_{F}' Q_{j-1}P_{j+1}+i\Omega_{F}' P_{j-1}Q_{j+1}\right)X_{j})
    \end{equation}
\end{widetext}
where $\Omega_{F}' = 1.2\Omega_{F}$ in our experiment. We then average over all the eigenstates of Eq.~(\ref{H_eff_odd}), projected into the odd configurational subspace, whose energies lie within a narrow window $[E_{0}-\Delta E, E_{0}+\Delta E]$ with equal weight. For a system initialized as a product state, $H_{A'}$ has no diagonal elements, thus the mean energy of the system $E_{0}$ is zero. The width of the energy window is chosen to be $\Delta E = 0.24\Omega_{F}$ \cite{rigol2008thermalization,kaufman2016quantum}. In Fig.~\ref{edfig:veri}, we show that the staggered magnetization resembles a smooth curve as a function of the eigenenergy. The total number of eigenstates within this energy window is 17, resulting in an averaged value of -0.019, which is plotted in Fig.~\ref{fig:KRT}(b). Correspondingly, for the odd subspace thermalizing from a different initial state as shown in Fig.~\ref{fig:subspace}(b), the statistically-averaged ground state probability is 0.53.

For the thermal predictions of the ground state probabilities in the even subspace [Fig.~\ref{fig:subspace}(a)], the total number of eigenstates within the energy window is 9 and the resulting ground state probability is 0.52.

\subsection{Staggered magnetization}
\label{Appendix:C.2}
In Fig.~\ref{fig:KRT}(b), we show the staggered magnetization for subarray $A'$ ($A$) to elucidate the dynamics under $H_{\text{KRT}}$ ($H_{\text{QPXPQ}}$). The full definitions for the staggered magnetizations are given below:
\begin{subequations}
\begin{eqnarray}
M_{A'}(t) &\equiv& \frac{1}{N_{A'}} \sum_{i=1}^{N_{A'}} (-1)^{i} \langle Z_{i}(t) \rangle \, , \\
M_{A}(t) &\equiv& \frac{1}{N_{A}} \sum_{i=2}^{N_{A'}-1} (-1)^{i} \langle Z_{i}(t) \rangle \, , 
\end{eqnarray}
\end{subequations}
where $i$ indexes the atom on odd sites of the entire chain and $N_{A} = N_{A'} - 2$.

\subsection{Local two-body observable}
\label{Appendix:C.3}
Besides the staggered magnetization, we use a local two-body observable to probe thermalization dynamics, defined as:
\begin{equation}
    \hat{O}(t) = \sum_{i,j \in A'} \langle \hat{n}_{i} \hat{n}_{j} \rangle 
\end{equation}
where $\hat{n}_{i}$ denotes the population in Rydberg state and $j = i + 1$ for the neighboring site \cite{kim2018detailed}. The pairs are chosen to reside only on subarray $A'$. We find that the long time behavior of the local two-body observable shows reasonable agreement with the expected value for a thermal state (see Fig.~\ref{edfig:veri}(b)).

\begin{figure}[t]
    \centering
    \includegraphics[keepaspectratio,width=8.8cm]{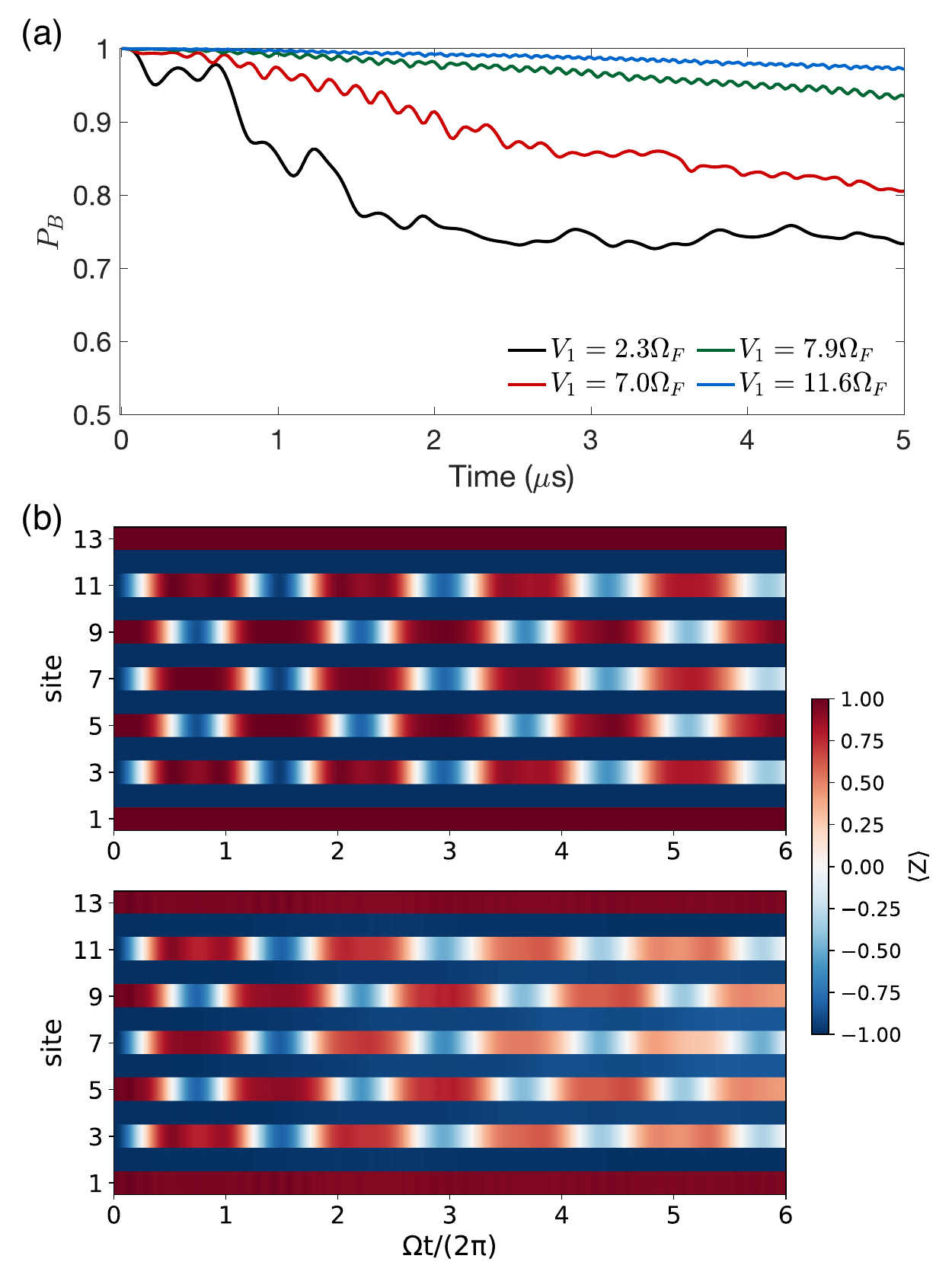}
    \caption{Deviations from effective models. (a) Effects of deviating from $H_{\text{KRT}}$. Starting from a $\mathbb{Z}_{4}$-ordered state with Rydberg excitations only on subarray $A$, atoms in subarray $B$ remain perfectly frozen in the ground state ($P_{B} = 1$) if $H_{\text{KRT}}$ is exactly realized. Off-resonant scattering on our experimental platform results in a PPXPP-type constraint that does not confine time-evolution to the odd or even subspace. Numerics calculated using Eq.~(\ref{eq:Ham_FFM}) with $\Omega = 2\pi\times1.48$~MHz show that smaller values of $V_{1}$ result in more serious deviations. Our experimental parameters correspond to $V_1 = 7.9\Omega_{F}$ (green curve), where relatively robust freezing of atoms in subarray $B$ is observed. (b) Comparison between the effective model $H_{\text{QPXPQ}}$ (top) and the full model $H_{\text{Ryd}}$ with $\Omega = 2\pi\times1.45$~MHz and $V_{1} = \Delta/2 = 2\pi \times 5$~MHz (bottom).}
    \label{edfig:model_dev}
\end{figure}

\section{Deviations and experimental imperfections}
\subsection{Deviations from effective Hamiltonians}
\label{Appendix:D.1}
Our ability to faithfully realize the effective Hamiltonian $H_{\text{KRT}}$ hinges on the assumption $V_{k-1}\gg\Omega_{F}\gg V_{k}$, where $V_{k} = V_{k-1}/2^{6}$ for our equally-spaced atom array. This assumption implies a trade-off between off-resonant scattering to undesired resonances (if $V_{k-1}$ is not large enough) and residual longer-range interactions (if $V_{k}$ is not small enough), both of which give rise to imperfections. In our experiments, we operate in the range $V_{k-1}/\Omega_{F} = 7.9 \pm 0.2$. Here we inspect how well imperfections are suppressed under our chosen experimental parameters.

We study the imperfections on $H_{\text{KRT}}$, focusing on off-resonant scattering. In $H_{\text{KRT}}$, we effect the multiple drives using Floquet frequency modulation. Compared to the ideal $H_{\text{KRT}}(k = 2)$ given by Eq.~(\ref{H_KRT}), there are additional off-resonant terms that would have been neglected from a rotating-wave approximation. To order $V_{1}$, these additional terms are:
\begin{widetext}
\begin{eqnarray}
    H_{\text{KRT}}^{(1)} = \frac{1}{2}\sum_{j=1}^{N}P_{j-1}\sigma^{j}_{+}P_{j+1}\,e^{-iV_{1}t} \times &&\left[-\Omega_{F} \left(Q_{j-2}P_{j+2}+P_{j-2}Q_{j+2}\right) \right.\nonumber \\
    && \left. + i\Omega_{F}' \left(1 + e^{2iV_{1}t}\right) P_{j-2}P_{j+2} + i\left(\Omega_{F}' e^{2iV_{1}t} - \Omega_{F}'' \right) Q_{j-2}Q_{j+2}\right] +\text{H.c.} \, ,  
\end{eqnarray}
\end{widetext}
where $\Omega_{F}'' = 0.20\Omega$ is the Rabi frequency corresponding to a detuning at $\Delta'' = 3V_{1}$. The first and last terms in the square brackets only lead to a slight adjustment in the Rabi frequencies and do not fundamentally alter the properties of the Hamiltonian. However, the middle term is a PPXPP-type term that does not, on its own, give rise to configurational subspaces within the first $V_{0}$ block. Therefore, we expect that if this term is significant, dynamics from an initial state in the odd (even) subspace will not be confined to the odd (even) subspace. We numerically probe this effect by measuring the ground state density of atoms on the even sites from an initial state in the odd subspace, as $V_{1}/\Omega_{F}$ is varied. From Fig.~\ref{edfig:model_dev}(a), we indeed see this leakage effect for smaller values of $V_{1}/\Omega_{F}$, where atoms on the odd sites do not remain frozen, whereas the leakage is well supressed for larger values. Importantly, our numerics show good supression of this leakage effect for our experimental parameters of $V_{1}/\Omega_{F} =7.9$. Longer-range interactions on our system are small: $V_{2} \approx 0.1\Omega_{F}$ and are not expected to be significant. In $H_{\text{QPXPQ}}$, there is only one off-resonant drive, detuned at $\Delta = 2V_{1}$. Thus, the leakage arising from a given finite $V_{1}/\Omega_{F}$ ratio occurs more slowly compared to that for $H_{\text{KRT}}$. 

A similar argument holds for $H_{\text{QPXPQ}}$, where we replace $\Omega_{F}$ with $\Omega$. The resulting scar dynamics from the effective $H_{\text{QPXPQ}}$ is compared against that from $H_{\text{Ryd}}$ in Fig.~\ref{edfig:model_dev}(b). The discrepancy between the full Rydberg and effective Hamiltonians contributes to an enhanced decay of the scar dynamics under $H_{\text{Ryd}}$ (see also Appendix~\ref{Appendix:E}).

\begin{figure}[t]
    \centering
    \includegraphics[keepaspectratio,width=8.8cm]{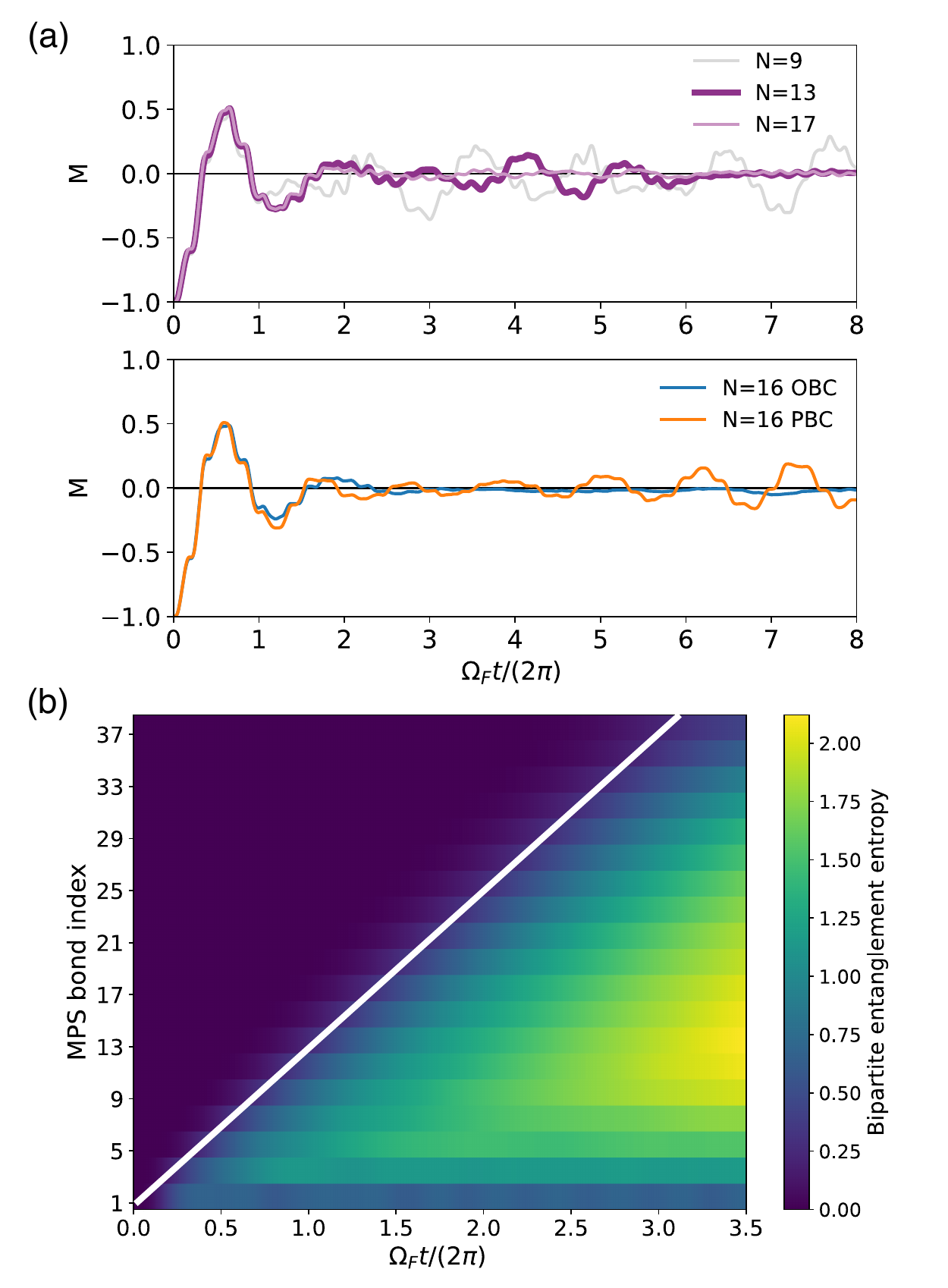}
    \caption{Analysis of finite size effects. (a) Time evolution of the staggered magnetization under $H_{\text{KRT}}$ with various system sizes (top) and with open boundary conditions versus periodic boundary conditions for a 16-atom chain (bottom). (b) Propagation of entanglement over time under $H_{\text{KRT}}$, calculated using TEBD methods. The MPS bond index $i$ partitions the chain into two subchains, one with atoms on sites $\{1, 2, \ldots, i\}$ and the other with the remaining atoms. The white line marks the boundary of the spreading as the light-cone limit. For both simulation, we use the same ratio of $\Omega_{F}'/\Omega_{F}$ = 1.2 as the one realized in our experiment.}
    \label{edfig:system_size}
\end{figure}

\subsection{Finite size effects}
In this section, we discuss the effect of finite array sizes and boundary conditions. Our system notably differs from others in which thermalization has been studied, since the kinetic constraints that are key to driving the dynamics are imposed \textit{locally}. Consequently, the boundaries in our model remain clean and minimally affect the dynamics. The main effect of the boundary atoms is to reduce the dimension of the Krylov subspace, which can be fully accounted for in simulations. 

To check the effect of the finite array size on Krylov-restricted thermalization, we compute the dynamics starting from a $\mathbb{Z}_{4}$-ordered initial state under $H_{\text{KRT}}$ (Eq.~(\ref{eq:H_KRT_OBC})) for various system sizes. Fig.~\ref{edfig:system_size}(a) shows that a quick decay of the magnetization towards the thermal value occurs even in small open chains, whereas the dimension of the relevant subspace primarily affects fluctuations around the thermal prediction. When periodic boundary conditions are considered, the translational symmetry further reduces the dimension of the Krylov subspace, giving rise to larger fluctuations (than for the corresponding array size with open boundary conditions) even at late times.

Since the effective interaction imposed through the kinetic constraints is short-ranged, the minimal evolution time required for a given non-equilibrium state to reach equilibrium could be related to the system size. It is thus informative to probe how entanglement would spread over time in such a system. We consider the dynamics of an open chain under $H_{\text{KRT}}$ and from an initial state where one excitation is located at the boundary (site 1). Using time-evolved block-decimation (TEBD) methods, we calculate the bipartite entanglement entropy associated with a given matrix product state (MPS) bond on the chain over time. Fig.~\ref{edfig:system_size}(b) shows that the entanglement propagates over 40 sites within the typical evolution time of our experiment (three Rabi cycles), where the velocity of propagation is consistent with the light-cone limit \cite{cheneau2012light}. Since 40 sites is more than three times the array size used in our experiment, the chosen evolution time should be sufficient for observing Krylov-restricted thermalization. 

\subsection{Effects of position disorder}
\label{Appendix:D.3}

\begin{figure}[t]
    \includegraphics[keepaspectratio,width=8.8cm]{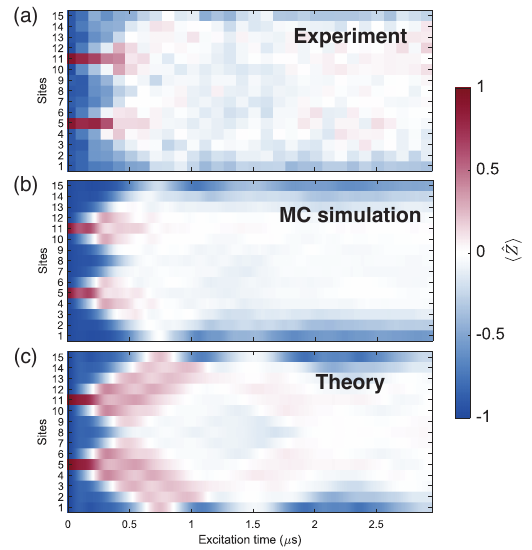}
    \caption{Dynamics resulting from two initial Rydberg excitations on a 15-atom chain, under the facilitation condition $\Delta = V_{0}$. (a) The excitations are experimentally observed to spread out, rather than be localized by position disorder, over time. There is a finite delay of 0.2~$\mu$s for the onset of spreading,  arising from the gap between the state preparation pulse and the main evolution pulse. (b) Monte Carlo simulation with position disorder but without single-atom decoherence. (c) Numerics for the ideal case, i.e.\ no disorder or decoherence. Both (b) and (c) use $H_{\text{Ryd}}$ with $\Omega=2\pi \times 1.37$~MHz and $V_{0}=2\pi \times 5$~MHz.}
    \label{edfig:2exc}
\end{figure}

To achieve facilitation, we detune our drive to match the van der Waals interaction strength. Since this interaction depends strongly on the distance between atoms, the dynamics under facilitation are expected to be highly sensitive to the position disorder of our atoms. In particular, a large enough position disorder may result in Anderson-like localization of the state in product-state space \cite{marcuzzi2017facilitation}. 

The specific metric governing whether or not localization occurs is the ratio between the interaction disorder and the drive strength, $\sigma_{V_0}/\Omega$. As reported in Reference \cite{marcuzzi2017facilitation}, localizing behaviour was observed at $\sigma_{V_0}/\Omega = 1$ whereas a ballistic spread of excitations was observed for $\sigma_{V_0}/\Omega \leq 0.19$. In our experiment ($T = 19(1)~\mu \text{K}$, trap depth of $0.6$~mK), we estimate the radial atomic position disorder to be $\sigma_r = 0.087~\mu$m. The resultant uncertainty in $V_{k-1}$ (where $k = 1$ for the facilitated QXQ model and $k = 2$ for the QPXPQ model) is $\sigma_{V_{k-1}} = 6|V_{k-1}| \sigma_r/(ka) = 2\pi \times 0.34$~MHz, where $a$ is the interatomic spacing of our chain. This yields a typical metric of $\sigma_{V_{k-1}}/\Omega = 0.24$, which lies near the threshold for ballistic spread of excitations. 

Experimentally, we characterize the spread of excitations in our setup to determine if the position disorder of atoms results in a significant localizing effect. We initialize our atom chain with two particular atoms in the Rydberg state, and then measure the site-resolved excitations under the facilitation condition $\Delta = V_{0}$ [Fig.~\ref{edfig:2exc}(a)]. We observe no significant localization of the excitations from position disorder, and the excitation spread dynamics are in good agreement with a Monte Carlo simulation that takes into account the same metric $\sigma_{V_{k-1}}/\Omega = 0.24$. 

\section{Quantum many-body scars}
\label{Appendix:E}

\begin{figure*}[t]
    \centering    
    \includegraphics[keepaspectratio,width=18cm]{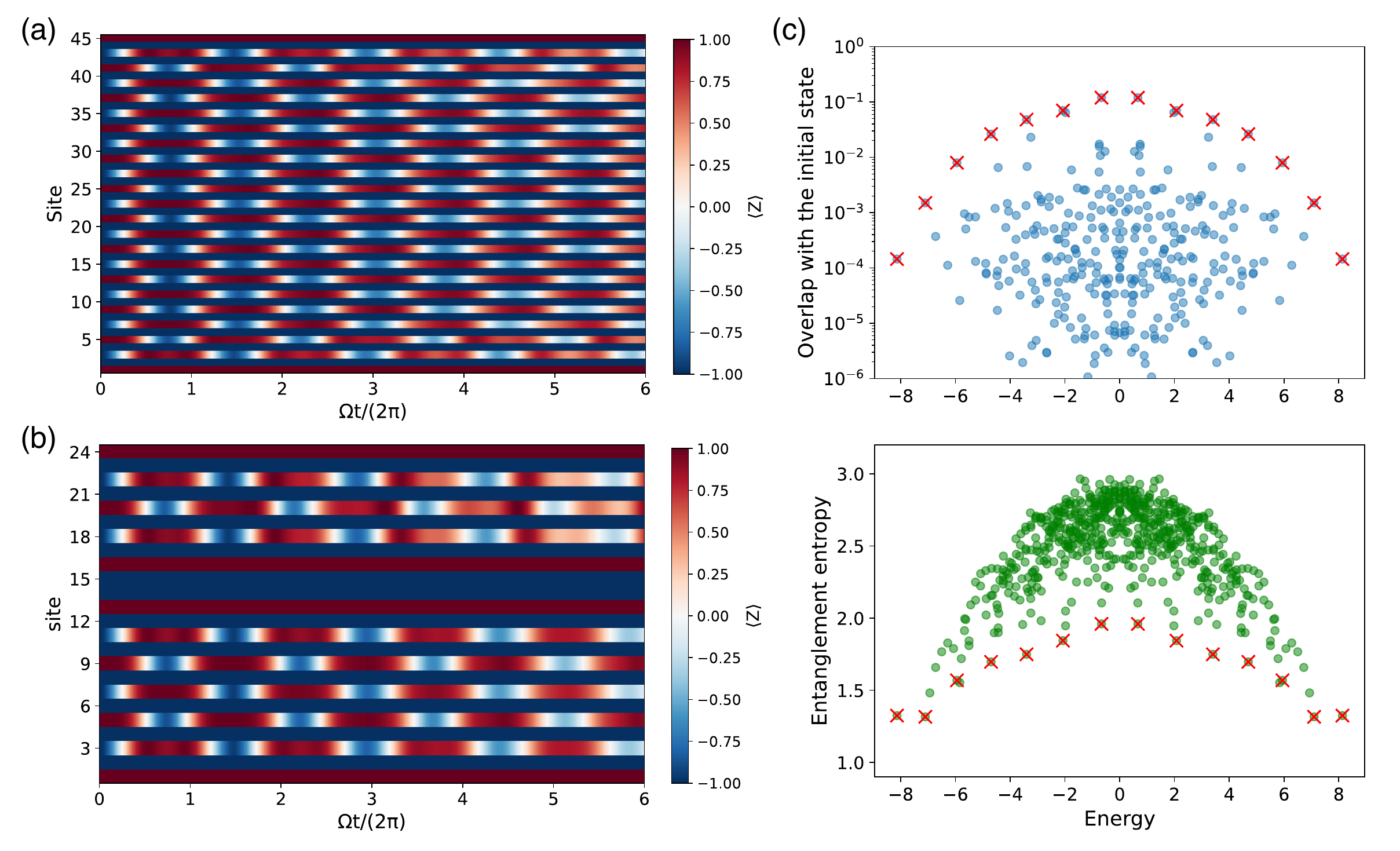}
    \caption{Quantum many-body scarring of the QPXPQ model. (a) Time evolution of single-site resolved $Z$ projections for a large system size ($N = 45$) using the TEBD technique applied to the effective Hamiltonian. (b) Scarring in the mixed subspace, calculated by exact diagonalization. A frozen substring disconnects the left and right parts of the chain. (c) Identification of scar states (marked as red crosses) as eigenstates that have low entanglement entropy (bottom) with roughly equal energy spacing. The same eigenstates have high overlap with the $\mathbb{Z}_{4}$-ordered initial state (top).}
    \label{edfig:Scar_M}
\end{figure*}

In the main text, we show quantum many-body scarring from an extended class of $\mathbb{Z}_{2k}$ initial states by analyzing the Fourier transform of the staggered magnetization of a subarray. However, given the relatively small subspace dimensions involved (e.g.\ 13 basis states for $H_{\text{QPXPQ}}$ in a 13-atom chain), one may be concerned that the observed oscillatory behavior may be attributed to transient dynamics in a reduced Hilbert space rather than quantum many-body scarring. To support our results, we present numerical simulations for a larger system with 45 atoms evolving under the QPXPQ Hamiltonian, calculated using the TEBD method. Fig.~\ref{edfig:Scar_M}(a) shows persistent $\mathbb{Z}_{4}$ scar oscillations despite having a subspace dimension that is two orders of magnitude larger than that used to demonstrate Krylov-restricted thermalization, indicating that the QPXPQ model indeed hosts quantum many-body scars.  

Besides the odd or even subspace, the mixed subspace can also host quantum many-body scars. As discussed in Appendix~\ref{Appendix:B.3}, a frozen substring $\ket{\cdots rggr \cdots}$ divides a given chain into two disconnected subchains, each beginning from a $\mathbb{Z}_{4}$-ordered initial state. Consequently, the time evolution of scar states can be understood as the combined revivals of these subchains, along with the frozen segment between them (Fig.~\ref{edfig:Scar_M}(b)).

Another critical metric for identifying scars is the entanglement entropy of eigenstates, where low-entanglement entropy eigenstates with approximately equal energy spacing correspond to atypical scar states \cite{turner2018weak,turner2018quantum}. To check for scar states in the QPXPQ model, we consider the odd configurational subspace ($n_{DW} = 0$) within the primary $V_{0}$ and $V_{1}$ block for a 29-atom open chain. We find eigenstates that are spaced roughly equally in energy and with characteristically low entanglement entropy, which we identify as scar states. These scar states display high overlap with the initial $\mathbb{Z}_{4}$-ordered state with Rydberg excitations on two ends of the chain, which accounts for the persistent oscillations in the staggered magnetization dynamics. 

We note that our effective models are of the PXP type, where the scars have an intrinsic lifetime $\tau_0$ due to approximate embedded algebra, finite overlap between the initial state and scar eigenstates, and  nearly-equal energy spacing between the scar eigenstates in a finite chain \cite{turner2018weak, turner2018quantum,choi2019emergent,lin2019exact,surace2021exact}. These reasons cause a similar decay of scars in the generalized QPXPQ models. Higher-order virtual couplings to states that affect the next-nearest-neighbor facilitation and the long-range tail of Rydberg interactions can contribute to additional decay of the scar, as described by the following empirical model (adapted from \cite{bluvstein2021controlling}):
\begin{equation}
\label{eq:scar_lifetime}
    \frac{1}{\tau} = \alpha (\frac{1}{2\pi}\sum\frac{\Omega^{2}}{4V_{1}})+\beta (\frac{1}{2\pi}\sum V_{2}) + \frac{1}{\tau_{0}}
\end{equation}
Eq.~(\ref{eq:scar_lifetime}) assumes no position disorder. On the other hand, our facilitation-enabled scars are sensitive to position disorder, which should yield a shorter observed scar lifetime.

\section{Numerical details}
\label{Appendix:F}

\begin{figure}[t]
    \centering
    \includegraphics[keepaspectratio,width=8.8cm]{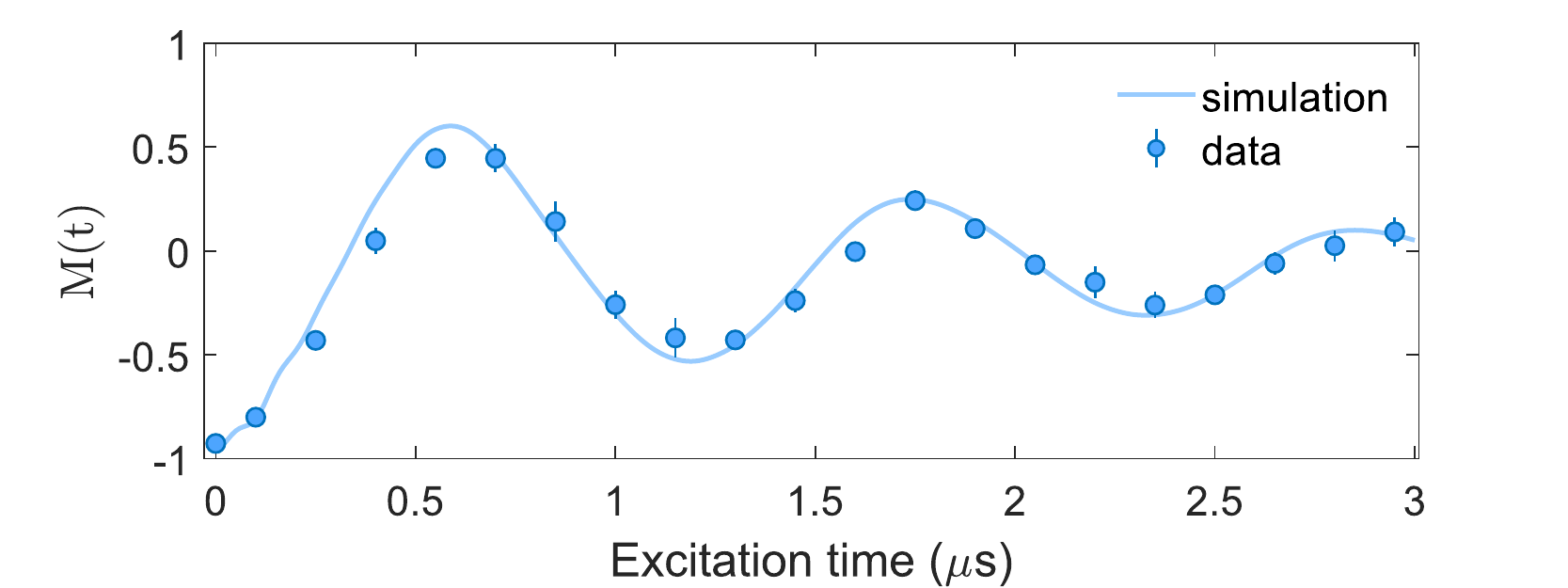}
    \caption{Dynamics of the PXP $\mathbb{Z}_{2}$ scar. Dependence of the staggered magnetization $M(t)$ versus time, corresponding to zero detuning. The solid curve is a theory simulation of $H_{\text{Ryd}}$ with $\Omega = 2\pi \times 1.38$~MHz, $V_0 = \Delta/2 = 2\pi \times 5$~MHz, and where the combined state preparation and measurement fidelity associated with the ground state is parameterized to be 0.99 and that for the Rydberg state is 0.96.
    }
    \label{edfig:stagmag}
\end{figure}

We use the ``Bloqade'', ``QuantumOptics'', and ``ITensors'' packages in Julia to generate the theory curves. For results in Figs.~\ref{fig:scars}(e),~\ref{fig:KRT}(b),~\ref{fig:KRT}(d),~\ref{edfig:model_dev}(a),~\ref{edfig:system_size}(a),~\ref{edfig:2exc}(b)(c),~\ref{edfig:Scar_M}(b) and ~\ref{edfig:stagmag}, we simulate $H_{\text{Ryd}}$ in Eq.~(\ref{eq:Hamiltonian}) and work in the complete Hilbert space of our system. For calculations in ~\ref{edfig:Scar_M}(a)(c), ~\ref{edfig:system_size}(b) we instead use the effective Hamiltonians.

For Figs.~\ref{fig:scars}(e),~\ref{fig:KRT}(b),~\ref{edfig:2exc}(b) and \ref{edfig:stagmag}, our numerics account for the atom position disorder, Doppler shift, and imperfect state preparation and detection fidelity. An example plot of the staggered magnetization dynamics at zero detuning is given in Fig.~\ref{edfig:stagmag}, which shows reasonable agreement between the experimental data and numerical simulations.  

In Fig.~\ref{fig:scars}(e), we perform a discrete Fourier transform on the measured staggered magnetization of subarray $A$. The intensity of the Fourier transform, denoted as $|S_{A}(\omega)|^{2}$, is normalized to the frequency domain $f$ of the single-sided spectrum. The frequency resolution is constrained by the total probe duration, which is 3~$\mu$s (2.4~$\mu$s) for the $\mathbb{Z}_{4}$ ($\mathbb{Z}_{6}$) experiment. The peak is then determined by identifying the maximum normalized intensity within the Fourier transform. Compared to the $\mathbb{Z}_{4}$ scar, the lower Fourier peak height of the $\mathbb{Z}_{6}$ scar primarily comes from the larger deviation of $H_{\text{Ryd}}$ from the effective QPPXPPQ Hamiltonian, where $V_{2}/\Omega = 3.0$. 

To generate the matrix plots in Figs.~\ref{fig:matrixplots},~\ref{fig:KRT}, and~\ref{edfig:ppxpp}, we use the ``QuTiP'' package \cite{johansson2012qutip} to simulate the effective Hamiltonians. To produce Fig.~\ref{edfig:veri}(a), we exactly diagonalize the effective Hamiltonian for a 7-atom open chain and calculate the eigenstates and the expectation values of $\hat{M}$. 


%

\end{document}